\documentclass[superscriptaddress,twocolumn,showpacs,amsmath,amssymb,nofootinbib]{revtex4}
\usepackage{graphicx}
\usepackage{dcolumn}
\usepackage{bm}

\begin{document}

\title{Energy dependence of nucleus-nucleus potential close 
to the Coulomb barrier}

\author{Kouhei Washiyama}
\email{washiyama@ganil.fr}
\affiliation{GANIL, CEA and IN2P3, Bo\^ite Postale 55027, 14076 Caen Cedex 5, France}
\affiliation{
Department of Physics, Tohoku University,
Sendai 980-8578, Japan}

\author{Denis Lacroix}
\affiliation{GANIL, CEA and IN2P3, Bo\^ite Postale 55027, 14076 Caen Cedex 5, France}


\date{\today}

\begin{abstract}

The nucleus-nucleus interaction potentials 
in heavy-ion fusion reactions  
are extracted from the microscopic time-dependent Hartree-Fock theory for 
mass symmetric reactions 
$^{16}$O${}+^{16}$O, $^{40}$Ca${}+^{40}$Ca, $^{48}$Ca${}+^{48}$Ca
and mass asymmetric reactions $^{16}$O$ +^{40,48}$Ca, $^{40}$Ca${}+^{48}$Ca, 
$^{16}$O$+^{208}$Pb, $^{40}$Ca+$^{90}$Zr.   
When the center-of-mass energy is much higher than the Coulomb barrier energy, potentials deduced 
with the microscopic theory identify with the frozen density approximation.
As the center-of-mass energy decreases and approaches the Coulomb barrier, potentials become energy 
dependent. This dependence signs dynamical reorganization of internal degrees of freedom
and leads to a reduction of the "apparent" barrier felt by the two nuclei during fusion 
of the order of $2-3 \%$ compared to the frozen density case. 
Several examples illustrate that the potential landscape changes rapidly 
when the center-of-mass energy is in the vicinity of the Coulomb barrier 
energy. The energy dependence is 
expected to have a significant role on fusion around the Coulomb barrier. 
\end{abstract}

\maketitle

\section{Introduction}

Heavy-ion fusion reactions give important information on dynamical 
evolution and dissipative phenomena in a quantum many-body system. 
Macroscopic models~\cite{norenberg76,hasse88,broglia91,reisdorf94,froebrich96} 
using suitable estimates of nucleus-nucleus 
potentials~\cite{bass,blocki77,myers00,krappe79,SL79,broglia91,moller94,ichikawa05},
and then coupled-channels theories~\cite{dasso83,BT98,nanda98}  
have been widely used 
to describe the entrance channel of fusion reactions. 
These models underline that the interplay 
between nuclear structure and dynamical effects is crucial to properly 
describe fusion reactions at energies close to the Coulomb barrier.
While in general rather successful, these methods have in common several 
drawbacks. First, nuclear structure and dynamical effects should be treated 
in a unified framework. Second, important effects should 
be guessed {\it a priori}. This has been illustrated recently 
to understand new high precision measurements at extreme 
sub-barrier energies~\cite{jiang02,jiang04,jiang05,jiang06,dasgupta07}, 
where different effects like incompressibility~\cite{misicu06}, 
nucleon exchange~\cite{esbensen07}, 
and the transition from di-nuclear to compound nucleus 
descriptions~\cite{ichikawa07}
have been invoked to understand experimental 
observations~\cite{jiang02,jiang04,jiang05,jiang06,dasgupta07}. 
Then, the hypothesis could only be checked {\it a posteriori}.
From this point of view, 
it is first desired to use theories where 
both nuclear structure and nuclear dynamics 
are considered in a unified framework. 
Second, the use of theories where all the physical 
effects mentioned above are automatically incorporated can be of 
particular interest to disentangle different contributions.  

Mean-field theories based on Skyrme Energy Density Functional (EDF)
provide a rather unique tool to describe nuclear structure 
and nuclear reactions over the whole nuclear chart. 
In nuclear reactions, application of the so-called 
Time-Dependent Hartree-Fock (TDHF), more than 30 years ago~\cite{bonche76,
koonin77,flocard78,bonche78,cusson78,koonin80,negele82},
to heavy-ion fusion reactions was a major step. 
With the increasing computer power, more and more 
accurate description of the nuclear reactions has been achieved. 
Most recent TDHF simulations include the spin-orbit force~\cite{kim97,
simenel01,denis02,simenel03,simenel07,simenel08}
and break all the symmetries 
(plane and axis symmetries generally assumed to speed up calculations). 
Moreover, all the terms of the Skyrme EDF used in nuclear structure can 
now be included~\cite{nakatsukasa05,maruhn05,umar05,umar06b,umar06,umar06c,
maruhn06,guo07}. 
The possibility to perform full three-dimensional calculations and 
to use effective forces consistent with nuclear structure 
is crucial to account for the richness of nuclear shapes 
that are accessed dynamically.
In addition, the great interest of dynamical mean-field theories 
with respect to other methods is that many effects which 
are known to affect fusion 
such as dynamical deformation, nucleon exchange, 
and nuclear incompressibility are automatically incorporated.     

The original purpose of the present work was to benchmark the method
proposed in Refs.~\cite{koonin80,denis02} 
to obtain nucleus-nucleus potential and one-body dissipation from the 
microscopic TDHF dynamics. The possibility to obtain such a potential 
from a mean-field theory has been studied in several works 
using the static EDF technique~\cite{denisov02,dobrowolski03,liu06,skalski07}. 
More recently,
a method called density-constrained TDHF (DC-TDHF)~\cite{umar85}, 
which combines TDHF dynamics with minimization technique 
under constraints on the one-body density, has been applied in 
Refs.~\cite{umar06,umar06c}, 
the latter being able to incorporate possible dynamical effects 
through the use of realistic density profiles obtained during the evolution.

Here, we consider a different approach based on a macroscopic reduction 
of the mean-field dynamics, called hereafter Dissipative-Dynamics TDHF 
(DD-TDHF). This technique could {\it a priori} give access not only 
to nucleus-nucleus potential but also to friction coefficients 
which play an important role in macroscopic 
models~\cite{gross74,blocki78,randrup80,randrup84} and 
has rarely been obtained from fully microscopic 
theories~\cite{koonin80,brink81}. 
The main difficulty of the macroscopic reduction 
is to guess the relevant collective degrees of freedom and their equation of motion.
Most models assume that the fusion problem can be reduced to 
a one-dimensional problem on the relative distance between nuclei. 
Here, we will test this hypothesis in TDHF 
and suppose that the dynamics is
described by a one-dimensional macroscopic dissipative dynamics. Since there is a freedom 
in the choice of macroscopic equations, the simple assumption made should first be 
validated. Thanks to alternative techniques used to infer Coulomb barriers 
from TDHF~\cite{umar06,simenel08}, we show that the DD-TDHF method 
can be a useful tool
to get precise information on potentials felt
during fusion. In particular, due to dynamical effects, 
the deduced nucleus-nucleus potentials depend explicitly on the center-of-mass 
energy close to the Coulomb barrier. 
This energy dependence, which was discussed in different models~\cite{seif06,feng08},
is studied in detail.

In this article, we concentrate on nuclear potentials study. Aspects related 
to dissipation will be discussed in Ref.~\cite{Kou08}.  
The paper is organized as follows.
Next section is devoted to the introduction of the DD-TDHF method.
In Sec.~III, we give the results and discussions on the extracted quantities.
A summary is given in Sec.~IV.

\section{Nucleus-nucleus potential from microscopic mean-field model}

In this section, the extraction of nucleus-nucleus interaction potentials from mean-field theories is discussed.
In macroscopic models, these potentials are generally displayed as 
a function of a few macroscopic collective degrees of freedom describing, 
e.g., the relative distance, shapes of nuclei, mass asymmetry. Here, we 
will concentrate on head-on collisions between initially spherical 
nuclei\footnote{Note that during the reaction nuclei might be deformed.} 
and assume that the collective space simply identifies with the relative 
distance $R$ between colliding nuclei. 
The validity of this approximation in the TDHF context will be discussed below.
In the following, we first present a general discussion on different methods 
to extract nucleus-nucleus potential from a mean-field theory.
 
\subsection{Some remarks on EDF}
 
The basic ingredient of a nuclear mean-field model is 
the energy functional of the one-body density~$\hat\rho$ 
denoted by ${\cal E} [\hat\rho]$.
In the nuclear context, ${\cal E} [\hat\rho]$ is expressed in terms of 
a few parameters generally related to the associated effective interaction. 
Here, we will use the Skyrme EDF with the SLy4d~\cite{kim97} parameters.
This choice is particularly suited to dynamical calculations
because of the removal of center-of-mass corrections in the fitting procedure 
of the force parameters~\cite{kim97}. In the EDF context, static 
properties of nuclei are deduced by minimizing the functional with respect to all possible one-body densities.
The great interest of EDF theory is that the initial complicated many-body 
problem is replaced by an independent particle problem. Indeed, 
the minimization procedure is equivalent to find the set of 
single-particle states that diagonalizes both the self-consistent mean-field
defined through the relation 
$h[\hat\rho]_{ij} = \partial {\cal E} [\hat\rho] / \partial \rho_{ji}$
and the one-body density. 
At the minimum, we have $[\hat h[{\hat\rho}],{\hat\rho}] = 0$.

\subsection{Illustration of fusion with time-dependent EDF}

The static EDF theory has also its dynamical counterpart, called time-dependent EDF\footnote{Although 
this theory has been called improperly TDHF in the past, we will continue to use this acronym in this work.} 
where the dynamical evolution of nuclear systems is replaced by the one-body density evolution, i.e.,
\begin{eqnarray}
i\hbar \frac{d\hat\rho}{dt}=[\hat h[\hat\rho],\hat\rho].
\end{eqnarray}   

\begin{figure}[tbhp]
\begin{center}\leavevmode
\includegraphics[width=\linewidth, clip]{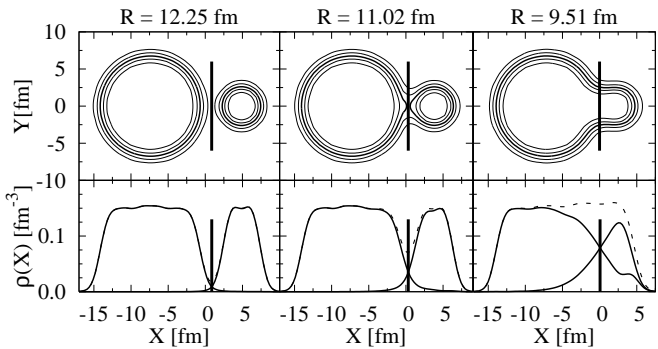}
\caption{Top: Density profiles~$\rho(X,Y,0)$
obtained with TDHF for the $^{16}$O${}+^{208}$Pb head-on collision at
$E_{\rm c.m.}=120$~MeV at three different relative distances.
The iso-densities (contour lines) are plotted at each 0.025~fm$^{-3}$.
The vertical lines indicate the positions of separation planes (see text).
Bottom: Total one-dimensional density~$\rho(X,0,0)$ (dashed line) obtained 
at the same relative distances. In each case, the two solid curves 
denote respectively $\rho_T(X,0,0)$ and $\rho_P(X,0,0)$ (see text). 
Again, the separation plane is presented by the vertical line. 
}
 \label{fig:densityopb}
\end{center}
\end{figure}

Dynamical calculations presented in this paper are performed with
the three-dimensional TDHF code developed by
P.~Bonche and coworkers with the SLy4d Skyrme effective force~\cite{kim97}.
As the initial conditions for the TDHF time evolution,
we prepare the density of colliding nuclei
by solving static HF equations with the same effective force 
as the one used in the TDHF. 
The step size in the coordinate space is 0.8~fm.
Then, we calculate the time evolution of the colliding nuclei
in the three-dimensional mesh.
The time step is 0.45~fm/c and the initial distance is set between
16~--~22.4~fm, depending on reaction.
We assume that 
the colliding nuclei follow the Rutherford trajectory
before they reach the initial distance for TDHF.
Thus, the initial positions and the momenta of the colliding 
nuclei are determined.
As an illustration, the density evolution of 
the $^{16}$O$+^{208}$Pb head-on collision
at center-of-mass energy $E_{\rm c.m.} = 120$~MeV are shown 
at three different relative distances in Fig.~\ref{fig:densityopb}. 

\subsection{Discussion of nucleus-nucleus potentials deduced from mean-field theories}

A good estimate of the interaction potentials felt by the two nuclei 
in the approaching phase within the EDF theory could be obtained 
assuming that the densities of the target and projectile
remain constant and equal to their respective ground state densities. 
This leads to the so-called {\it Frozen Density} (FD) approximation 
(see for instance Ref.~\cite{denisov02}). 
In this limit, the interaction potential between 
a target and projectile with their ground state densities respectively denoted by $\hat\rho_T$ and $\hat\rho_P$ reads
\begin{eqnarray}
V^{FD}(R) = {\cal E}[\hat\rho_{P+T}](R) - {\cal E}[\hat\rho_T] - {\cal E}[\hat\rho_P], 
\label{eq:frozen}
\end{eqnarray}
where ${\hat \rho}_{P+T}={\hat \rho}_{P}+{\hat \rho}_{T}$ 
is the total density obtained by summing 
the densities of the target and projectile assuming that
their centers of mass are at a given relative distance~$R$.
Note that ${\cal E}[\hat\rho_{P+T}]$ here neglects the Pauli effect
by the two overlapping densities. As we will see the following, 
we use the property of the extracted potentials with FD technique 
essentially at the Coulomb barrier. 
At this point, overlap of the two densities is small and Pauli effect 
is expected to be accordingly small. It should however be kept in mind 
that a proper account of the Pauli principle would lead to an increase of 
the potential for small relative distances.
    
\subsection{Matching TDHF with a two-body collision problem}

The FD approximation is expected to break down if strong reorganization 
of internal one-body degrees of freedom occurs in the approaching phase.       
In the following, methods to extract nucleus-nucleus potential directly from 
TDHF without assuming frozen densities are discussed. 

\subsubsection{Definition of the separation plane}

The first step is to properly define  
the collective coordinate $R$ that separates the two sub-systems.
Here, we follow macroscopic models and define the plane of separation at the neck 
position. In practice, the neck position is obtained by considering the two densities
\begin{eqnarray}
\rho_{T/P}({\bf r},t) = \sum_{n\in  {T/P}} |\varphi^{T/P}_n(\mathbf{r} ,t)|^2, \nonumber
\end{eqnarray} 
{where $\varphi^{T}_n(\mathbf{r},t)$ (resp. $\varphi^{P}_n(\mathbf{r},t)$)
denote single-particle wave functions}
initially in the target (resp. projectile) which are propagated 
in the mean-field of the composite system up to time~$t$. 
Note that, if the approaching phase is fast enough, these densities have 
no time to evolve and is expected to be very close to their respective ground state densities.
The separation plane at a given time~$t$ 
is then defined at the position where iso-contours of the two densities 
$\rho_{T}({\bf r},t)$ and $\rho_P({\bf r},t)$ cross.
An example of densities $\rho_{P,T}({\bf r})$ (solid lines) as well as 
the deduced separation plane (vertical thick line) is given for the reaction 
$^{16}$O${}+^{208}$Pb in bottom of Fig.~\ref{fig:densityopb}. 
This figure illustrates that the separation plane corresponds to 
the geometrical neck as generally defined in leptodermous systems.
 
\subsubsection{Two-body kinematics}

Once the separation plane is defined, all quantities relative to the dynamics 
of the two sub-systems can be calculated.
We associate to each sub-space an index~$i=1,2$ and
a density $\hat\rho_i(t)$, which equals the total density in the sub-space "$i$" and cancels 
out in the opposite side of the separation plane. 
Then, all quantities of interest could be computed like the number of nucleons 
in each side of the separation plane, i.e., $A_i(t)\equiv {\rm Tr}(\hat\rho_i(t))$.
Since we are considering head-on collisions along x-axis, the
center-of-mass coordinate $R_i(t)$ and associated momentum $P_i(t)$ express as
\begin{eqnarray}
R_i(t) \equiv {\rm Tr}(\hat{x}\hat\rho_i(t))/A_i(t), \hspace{1em} 
P_i(t) \equiv {\rm Tr}(\hat{p_x}\hat\rho_i(t)).
\end{eqnarray}
We can also compute the inertial mass of the two sub-systems, denoted by $m_i$,
from the TDHF evolution using $m_i=P_i/\dot{R_i}$.
Once these quantities are obtained, the TDHF dynamics can be reduced to 
the two-body collision problem where the relative distance $R(t)=R_1-R_2$, 
associated momentum $P(t)=(m_2P_1-m_1P_2)/(m_1+m_2)$,
and reduced mass 
\begin{eqnarray}
\mu =\frac{m_1 m_2}{m_1 +m_2}
\label{eq:mu}
\end{eqnarray}
are computed at each time step. Note that, when the two nuclei are far from 
each other, the reduced mass properly identifies with its initial value 
$\mu_{\rm ini} = m A_T A_P/(A_T+A_P)$, where $A_T$ and $A_P$ denote 
respectively the initial target and projectile mass\footnote{The 
discrepancy on the reduced mass found here very close to $\mu(R)/\mu_{\rm ini} \simeq 1$
and the one around $0.67$ in Ref.~\cite{denis02} 
was due to time-odd terms of the Skyrme field which was not properly computed 
in the original version of 3D code~\cite{kim97}. This problem is now fixed.}
and $m$ is the nucleon mass.
It is worth mentioning that, while for symmetric reactions $\mu$ remains 
constant during the collision, for asymmetric case $\mu$ may vary 
after the contact, i.e., $\mu = \mu(R)$. 
The critical discussion on the possible influence of 
this variation on extracted potentials is presented in Sec.~\ref{section:mu}.

\subsection{Dynamical effects on nucleus-nucleus potentials}
The violation of the FD prescription might at least be assigned to two effects:
{\bf (i)} In TDHF, the total one-body density $\hat\rho(t)$ does evolve in time. 
Therefore, the frozen density used in Eq.~(\ref{eq:frozen}) should 
{\it a priori} be replaced by the density reached dynamically 
during the evolution to get a more realistic  
nucleus-nucleus potential from TDHF. {\bf (ii)} A second crucial aspect
is that part of the relative kinetic energy is transformed into internal excitations. 
This effect is generally treated as a dissipative process in macroscopic theories. 

In order to include the effect {\bf (i)} and extract interaction potentials 
which account for possible evolution of the density along the TDHF path, 
the so-called density-constrained TDHF (DC-TDHF) technique has been 
developed~\cite{umar85,umar06,umar06c}. In this method, at each time-step, 
$\rho({\bf r},t)$ is deduced from TDHF. Then, the EDF is minimized 
under the constraint that the total density matches $\rho ({\bf r},t)$. 
Denoting the minimized energy by ${\cal E}_{DC}[\hat\rho(t)](R)$,
the potential is then given by
\begin{eqnarray}
V^{DC}(R) = {\cal E}_{DC}[\hat\rho(t)](R) - {\cal E}[\hat\rho_T] - {\cal E}[\hat\rho_P].
\end{eqnarray}
The great interest of the DC-TDHF method lies in the possibility to access the adiabatic potential 
accounting for realistic density profiles.

\subsection{Matching TDHF with binary dissipative collisions}

An alternative technique proposed in Ref.~\cite{koonin80}, 
preliminary tested in Ref.~\cite{denis02}, consists in 
assuming that the time evolution of $R$ and its canonical momentum $P$ 
obey a classical equation of motion
including a friction term which depends on the velocity $\dot{R}$:
\begin{eqnarray}
\frac{dR}{dt}&=&\frac{P}{\mu (R)},\nonumber\\
\frac{dP}{dt}&=&-\frac{dV}{dR}^{DD}+\frac{1}{2}\frac{d\mu(R)}{dR} \dot{R}^2
                -\gamma (R)\dot{R},
\label{newtonequation}
\end{eqnarray}
where $V^{DD}(R)$ and $\gamma (R)$ denote respectively the nucleus-nucleus potential 
and friction coefficient (here "DD" stands for Dissipative Dynamics). 
The friction coefficient $\gamma (R)$ describes the effect of energy dissipation 
from the macroscopic degrees of freedom to the microscopic ones.
The great interest of this method 
is the possibility to access interaction potentials which account for possible dynamical 
effects and to get information on dissipative process from a fully microscopic theory.
In the following, we show that this method is a valuable tool.

\subsubsection{Discussion on the R-dependent mass}

Before presenting applications, additional remarks on the origin of the $d\mu(R)/dR$ term
in Eq.~(\ref{newtonequation}) are mandatory. To obtain the equation, 
we implicitly assumed that the total energy can be written as a sum of 
a Hamiltonian part and a dissipative part $E_{\rm diss}$, i.e. 
\begin{eqnarray}
E = \frac{P^2}{2 \mu (R)} + V^{DD}(R) + E_{\rm diss},
\end{eqnarray}
where the reduced mass depends explicitly on $R$. 
For canonical variables $(R,P)$, the Hamilton part in Eq. (\ref{newtonequation}) 
arises from the derivative of the kinetic and $V^{DD}$ part in previous equation.
Using  $\frac{1}{2}\frac{d\mu(R)}{dR} \dot{R}^2 = -\frac{d}{dR} \left(\frac{P^2}{2\mu(R)}\right)$, 
we finally obtain the second equation in~(\ref{newtonequation}). 
The appearance of the $R$-dependent reduced mass in dynamical mean-field 
calculations has been reported in Ref.~\cite{umar06c,denis02} and turned out to be weak 
before the Coulomb barrier. In all applications presented below we explicitly neglected the
second term in the evolution of $P$. Indeed, including it or not or taking directly 
$\mu(R) = \mu_{\rm ini}$ has no effect on the barrier height and position presented in this work. 
For the sake of completeness this aspect is illustrated in Sec.~\ref{section:mu}.

\section{Application}

\subsection{Procedure to obtain $V^{DD}(R)$ and $\gamma(R)$}
\label{section:procedure}

In Eq. (\ref{newtonequation}), we have three unknown quantities: the reduced mass $\mu(R)$ entering in the 
first equation, the friction coefficient $\gamma(R)$ and the nucleus-nucleus potential $V^{DD}(R)$ appearing in the relative 
momentum evolution.
The reduced mass can be directly deduced from a single TDHF trajectory. Indeed, both $R(t)$ and $P(t)$ can be 
computed at all time along the mean-field trajectory. From $R(t)$, $\dot{R}(t)$ is simply computed from
\begin{eqnarray}
\dot{R}(t) \equiv \frac{R(t+\Delta t) - R(t-\Delta t)}{ 2 \Delta t}
\end{eqnarray}
where $\Delta t$ corresponds to the numerical time step used in TDHF. We finally deduced $\mu(R)$ from $P(t)/\dot{R}(t)$.

To obtain the two remaining unknown quantities $\gamma(R)$ and $V^{DD}(R)$, a single TDHF trajectory is not sufficient.
We assume that  
the potential energy and the friction parameter are not affected by
a slight change in center-of-mass energy and consider two head-on collisions
with energies $E_{I} = { E}_{\rm c.m.}$ and $E_{II} = {E}_{\rm c.m.} + \Delta E$ 
(in practice $ \Delta E/{ E}_{\rm c.m.} \simeq 1-2\%$ is used). A couple of canonical variables 
$(R_{I/II},P_{I/II})$ is associated to each trajectory. 
Assuming that Eq.~(\ref{newtonequation}) applies in both cases with the same potential and 
friction, we deduce 
\begin{eqnarray}
\gamma(R) = -\frac{[\dot P_I]_{R_I = R} - [\dot P_{II}]_{R_{II} = R}}
{[\dot R_I]_{R_I = R} - [\dot R_{II}]_{R_{II}=R}}.
\end{eqnarray}  
Then, using one of the trajectories, we obtain $dV^{DD}/dR$ as a function of relative distance.
The potential $V^{DD}(R)$ is deduced by integration over $R$ using its asymptotic Coulomb potential 
at large relative distances. The present method clearly relies on the hypothesis that the mean-field dynamics could 
properly be reduced to a one-dimensional macroscopic description. As we will see, this 
potential compares rather well with other techniques validating {\it a posteriori}
the macroscopic reduction used in this work. Finally, $V^{DD}(R)$ is also expected to contain
dynamical effects like density evolution.

\subsection{Illustrative example: $^{16}$O$+^{16}$O}

\begin{figure}[tbhp]
\begin{center}\leavevmode
\includegraphics[width=\linewidth, clip]{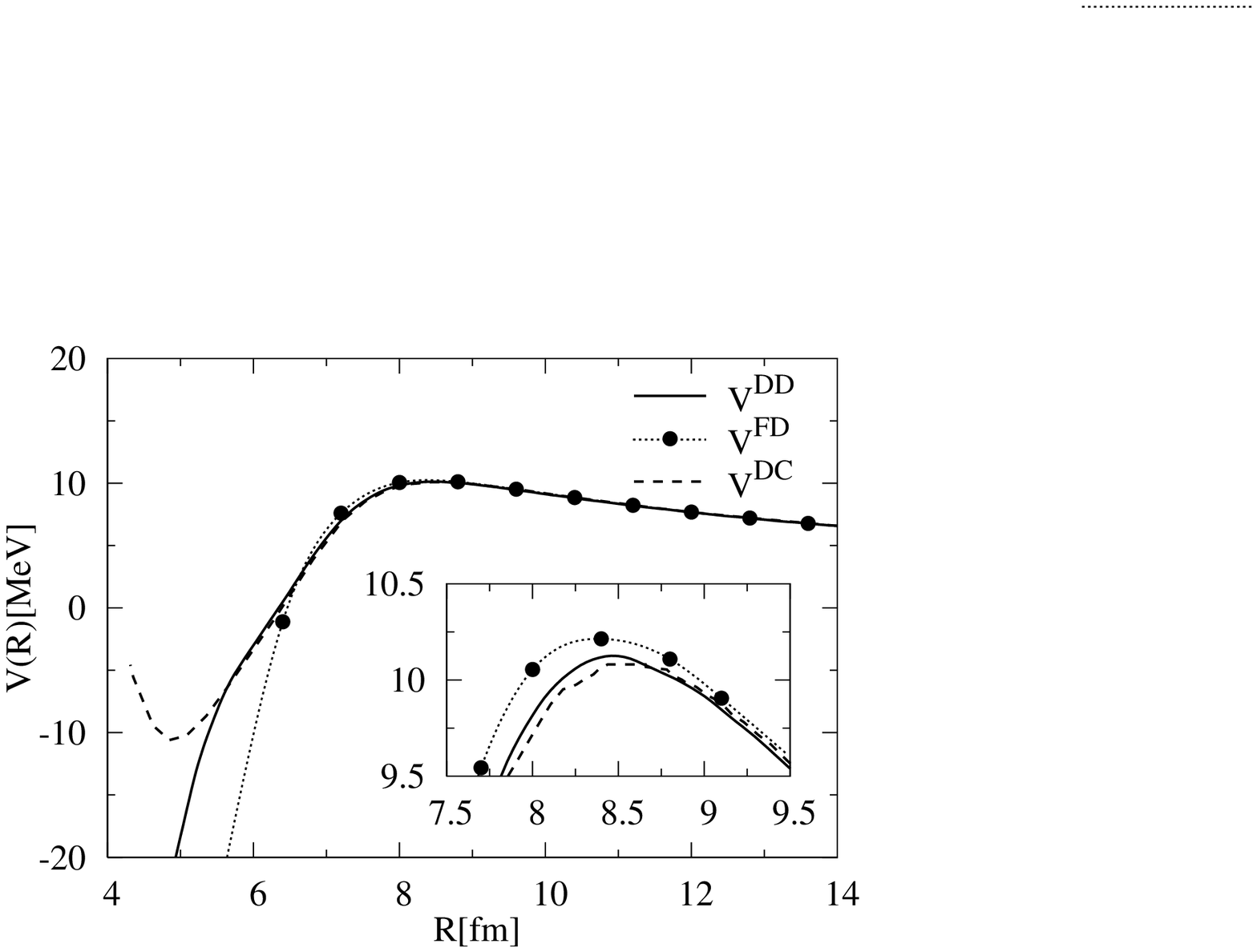}
\caption{
Comparison of potential energies for the $^{16}$O$+^{16}$O reaction
obtained from different models.
The solid, dashed, and filled circles-dotted line correspond to the DD-TDHF, 
DC-TDHF~\cite{umar06}, and FD potentials, respectively.
A zoom on the Coulomb barrier region is also shown in the insert. 
}
 \label{fig:poto16o16}
\end{center}
\end{figure}

\begin{figure}[tbhp]
\begin{center}\leavevmode
\includegraphics[width=\linewidth, clip]{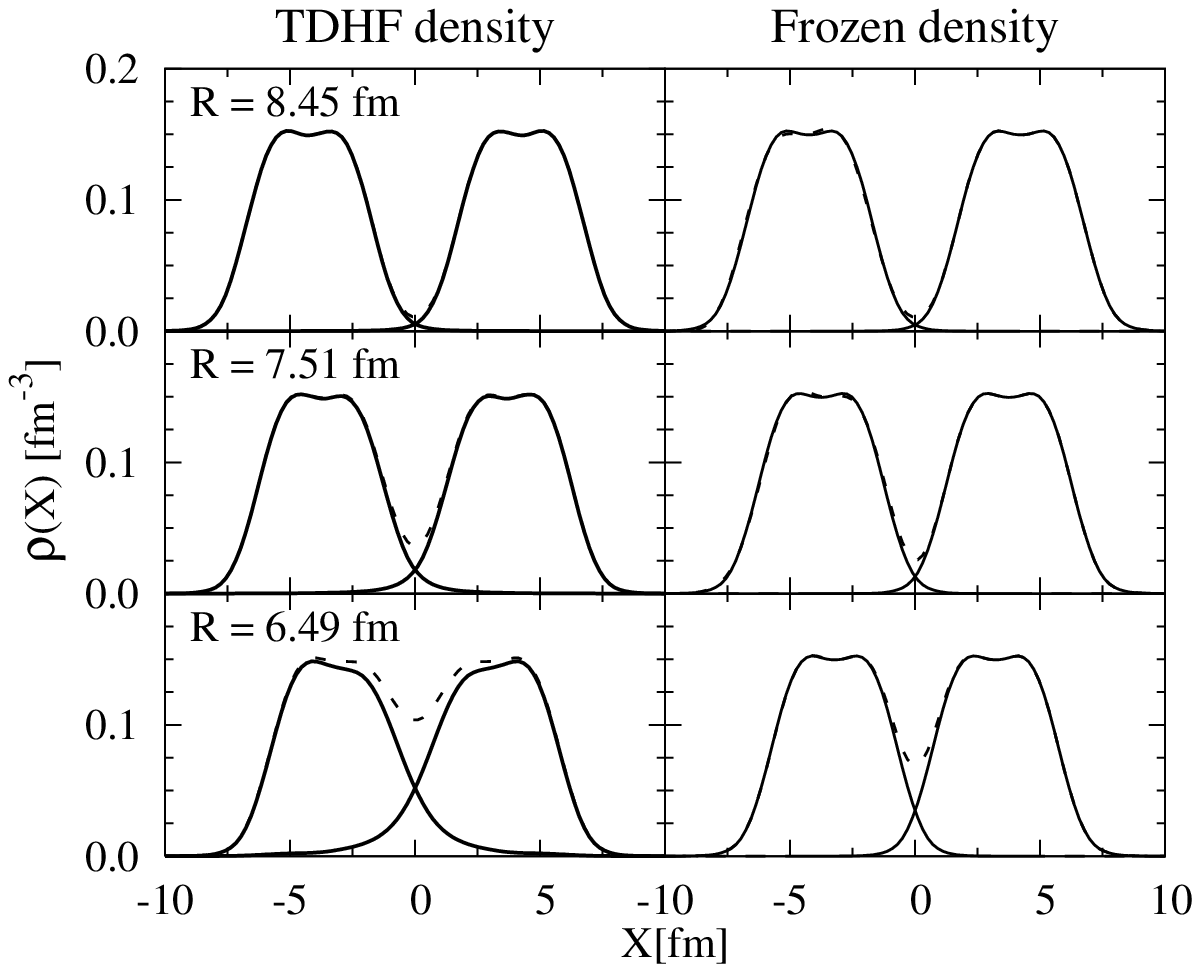}
\caption{Left:
Density profiles $\rho(X,0,0)$ (dashed) and $\rho_{P/T}(t)$ (solid lines) obtained with TDHF
for the head-on $^{16}$O${}+^{16}$O collision
at $E_{\rm c.m.}=34$~MeV at three different times.
Each value of relative distance $R$ is indicated 
in each left panel. Right: Densities $\rho(X,0,0)$ (dashed lines) 
obtained at the same relative distances within the FD approximation are shown. 
In this case, $\rho_{P/T}(X,0,0)$ (solid lines) identify with the ground state densities of the $^{16}$O nucleus.
}
 \label{fig:densityo16o16}
\end{center}
\end{figure}

The potential $V^{DD}(R)$ obtained with the dissipative dynamics reduction 
method for the $^{16}$O${}+^{16}$O reaction at  $E_{\rm c.m.}=34$~MeV 
(and $\Delta E = 1$ MeV) 
is displayed (solid line) in Fig~\ref{fig:poto16o16}.
The DC-TDHF (dashed line) potential obtained in 
Ref.~\cite{umar06}\footnote{Note, however, that a different set of parameters 
was used for the Skyrme effective interaction and numerical aspects.} 
and the FD potential (filled circles-dotted line) 
are also displayed for comparison.    
Figure~\ref{fig:poto16o16} shows that potentials extracted from 
the DD- and DC-TDHF methods are very close from each other (almost identical) 
even well inside the Coulomb barrier (up to $R = 5.3$~fm). 
The fact that the potential 
deduced from our method matches the DC-TDHF result gives confidence in the specific macroscopic 
equation (Eq.~(\ref{newtonequation})) retained to reduce the microscopic dynamics. 
In addition, both methods are almost identical to the FD 
description (for $R \ge 6.5$~fm). 
This indicates that little reorganization of densities
occurs in the approaching phase. This is indeed confirmed
in Fig.~\ref{fig:densityo16o16} 
where the TDHF density profiles obtained at different relative distances (Left) are directly compared 
to densities used in the FD approximation for the same $R$ (Right).  
At and below the estimated barrier radius $R_B= 8.46$~fm,
little difference between the densities $\rho_{P/T}(X,0,0)$ (Left-solid line) and the ground state densities (Right-solid lines)
can be seen. As a consequence, the Coulomb barrier predicted 
by TDHF is almost identical to the one obtained in the FD 
case (the difference being less than 0.1~MeV).
It is worth mentioning at that point that our method
assumes neither sudden nor adiabatic approximation. Last, another conclusion 
that could be drawn from the matching between DD-TDHF or DC-TDHF and the FD approximation is that 
Pauli blocking effects which are automatically incorporated in the two former approaches and partially neglected
in $V^{FD}$ do not seem to play a significant role close to the Coulomb barrier in $^{16}$O${}+^{16}$O.

Figure~\ref{fig:poto16o16} indicates that dynamical effects affect marginally the potential felt 
by the two partners in the approaching phase. In the next section, we will indeed see that similar conclusions 
hold in most cases studied when $E_{\rm c.m.}$ is well above the Coulomb barrier. This is the case presented here 
where $E_{\rm c.m.}$ is three times more than the Coulomb barrier. In this limit, the spatial organization of 
nucleons inside each nucleus is almost frozen before the contact.

\subsection{Systematic study of nucleus-nucleus potential at high center-of-mass energy}

Previous study is extended to fusion reactions with 
various combinations of nuclei.
Since deformed nuclei have  
orientations with respect to the collision axis,
which increases macroscopic degrees of freedom to be considered,
we concentrate on collisions involving spherical nuclei.
The DD-TDHF technique is applied to the systems $^{40}$Ca${}+^{40}$Ca, 
$^{48}$Ca${}+^{48}$Ca for mass symmetric reactions and
$^{16}$O$ +^{40,48}$Ca, $^{40}$Ca${}+^{48}$Ca, $^{16}$O$+^{208}$Pb, 
$^{40}$Ca$+^{90}$Zr for mass asymmetric reactions.

\begin{figure}[tbhp]
\begin{center}\leavevmode
\includegraphics[width=\linewidth, clip]{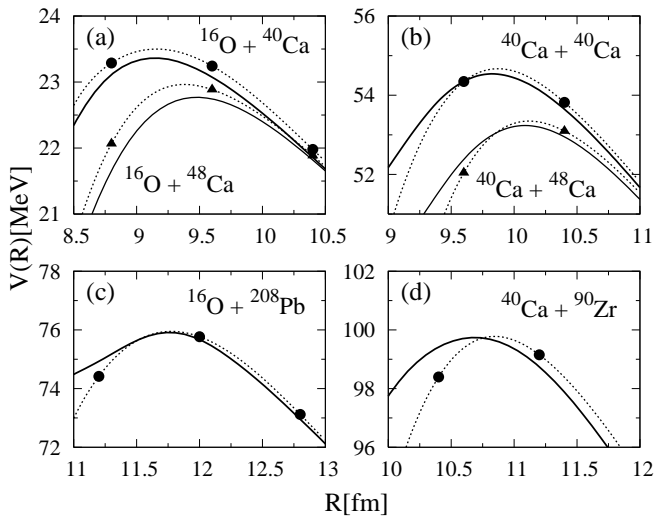}
\caption{
Potentials extracted from the DD-TDHF (solid lines) 
at high center-of-mass energies compared 
to $V^{FD}$ (filled circles- and triangles-dotted lines).
}
\label{fig:potall}
\end{center}
\end{figure}

\begin{figure}[tbhp]
\begin{center}\leavevmode
\includegraphics[width=0.85\linewidth, clip]{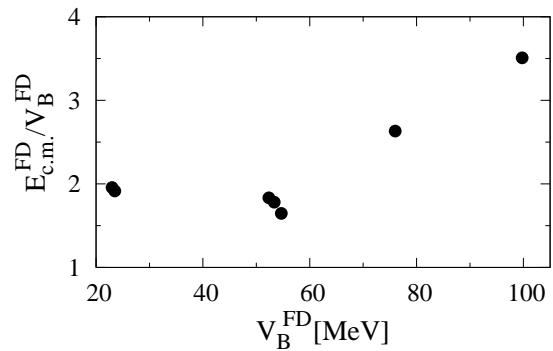}
\caption{Minimal center-of-mass energy, denoted by $E^{FD}_{\rm c. m.}$, 
for which the potential deduced 
from the DD-TDHF method identify with the FD case. 
This quantity is presented as a function of the FD barrier energy.  
}
\label{fig:ecmall}
\end{center}
\end{figure} 

In all systems, we do expect that, when center-of-mass energy increases, 
the potential will identify with the FD case. We have checked that 
this is indeed the case and 
identifies the minimum energy for which the FD limit is reached. 
Potentials obtained with the DD-TDHF are displayed 
by the solid lines in Fig.~\ref{fig:potall} and are systematically compared 
with the FD approximation. In Figs.~\ref{fig:potall}(a) and \ref{fig:potall}(b), 
the center-of-mass energy used to perform the macroscopic reduction 
is about two times the Coulomb barrier energy.
Similarly to the $^{16}$O${}+^{16}$O reaction case, all the 
examples presented in Figs.~\ref{fig:potall}(a) and \ref{fig:potall}(b) 
follow closely the FD approximation. 
Higher center-of-mass energies have to be used to reach the FD case in the 
systems presented in Figs.~\ref{fig:potall}(c) and \ref{fig:potall}(d). 
We report in Fig.~\ref{fig:ecmall} the center-of-mass 
energy threshold, denoted by $E^{FD}_{\rm c.m.}$, 
which corresponds to the minimal $E_{\rm c.m.}$ for which the DD-TDHF method gives the FD results 
(within $5\%$ in general)
as a function of the FD barrier. 
The different Coulomb barrier energies deduced from the DD-TDHF method applied at high center-of-mass 
energies (denoted by $V^{DD}_{B}$ (high $E_{\rm c.m.}$)) are reported 
in Table~\ref{tab:vbrb}. These barriers are systematically compared with the FD
case and experimental data taken from Refs.~\cite{nanda98,vaz81,newton04}.
Overall, we see that the DD-TDHF method applied at high center-of-mass energy 
gives a qualitative agreement with experiments. It is however noticeable 
that the barrier height is systematically higher than the experimental 
observation and that the discrepancy increases as $Z_PZ_T$ increases.

We will see in the following that part of the difference observed could 
be understood in terms of departure from the FD limit as 
the center-of-mass energy approaches the Coulomb barrier. Indeed, as the 
energy decreases, densities have more time to reorganize.

\begin{table*}[bthp]
\caption{\label{tab:vbrb}
Energy and radii of the Coulomb barrier extracted from the DD-TDHF
method. Here, $V^{DD}_B$ (high $E_{\rm c.m.}$) refers to barrier deduced 
for $E_{\rm c.m.} > E^{FD}_{\rm c.m.}$ 
while $V^{DD}_B$ (low $E_{\rm c.m.}$) corresponds to the lowest Coulomb barrier
deduced from TDHF using $E_{\rm c.m.} \simeq V^{FD}_B$. 
The experimental values taken from Refs.~\cite{vaz81,nanda98,newton04}
are reported when available.}
\begin{ruledtabular}
\begin{tabular}{c|cccccccc}
~Reaction~&$V_B^{FD} $ (MeV)&$V_B^{DD} $ (MeV)
&$V_B^{DD} $ (MeV)&$V_B^{\rm exp}  $ (MeV)
&$R_B^{FD}$ (fm) &$R_B^{DD}$ (fm)
&$R_B^{DD}$ (fm) &$R_B^{\rm exp}$ (fm)\\
&&(high $E_{\rm c.m.}$)&(low $E_{\rm c.m.}$)&&
&(high $E_{\rm c.m.}$)&(low $E_{\rm c.m.}$)& \\
\hline \hline
$^{16}$O$ +^{16}$O  & 10.2 & 10.13 & 10.12 & 10.61~\cite{vaz81}
                    & 8.4   & 8.46  &  8.52 &  7.91~\cite{vaz81}\\
$^{16}$O$ +^{40}$Ca & 23.5  & 23.36 & 23.07 & 23.06~\cite{vaz81}
                    & 9.2   &  9.18 &  9.50 &  9.21~\cite{vaz81}\\
$^{16}$O$ +^{48}$Ca & 23.0  & 22.77 & 22.48 &  
                    &  9.4  &  9.50 &  9.75 & \\
$^{40}$Ca$+^{40}$Ca & 54.7 & 54.54 & 53.35 & 52.8~\cite{nanda98} 
                    &  9.8  &  9.82 & 10.32 & \\
$^{40}$Ca$+^{48}$Ca & 53.4 & 53.24 & 52.13 & 52.00~\cite{newton04}
                    & 10.1  & 10.09 & 10.56 &  9.99~\cite{newton04}\\
$^{48}$Ca$+^{48}$Ca & 52.4 & 52.13 & 50.97 & 51.49~\cite{newton04}
                    & 10.3  & 10.38 & 10.82 & 10.16~\cite{newton04}\\
$^{16}$O$+^{208}$Pb & 76.0  & 75.91 & 74.51 & 74.52~\cite{newton04}
                    & 11.8 & 11.74 & 12.14 & 11.31~\cite{newton04}\\
$^{40}$Ca$+^{90}$Zr & 99.8 & 99.98 & 97.71 & 96.88~\cite{newton04}
                    & 10.8  & 10.63 & 11.27 & 10.53~\cite{newton04}\\
\end{tabular}
\end{ruledtabular}
\end{table*}

\subsection{Center-of-mass energy dependence of nucleus-nucleus potential close to the Coulomb barrier}

In the previous examples, we have 
determined 
the typical center-of-mass energy above which the potential deduced from 
the DD-TDHF technique corresponds to the FD case. 
Here, we show that the extracted potential energy 
is slightly modified as the center-of-mass energy decreases and approaches the Coulomb barrier. 
As shown below, this energy dependence of the nucleus-nucleus potentials underlines the role of dynamical effects.     

\subsubsection{Dynamical reduction of the Coulomb barrier energy}
\begin{figure}[tbhp]
\begin{center}\leavevmode
\includegraphics[width=0.9\linewidth, clip]{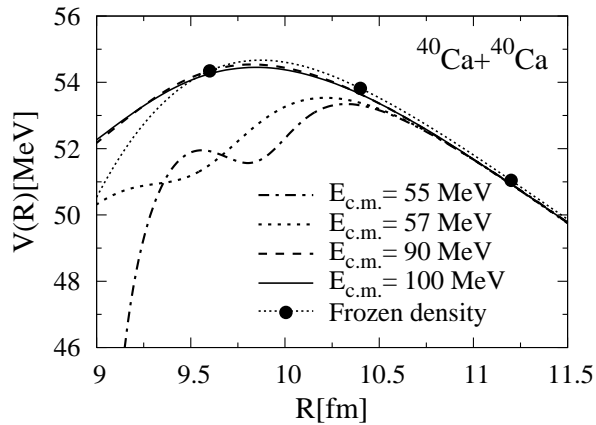}
\caption{
Potential energy for the $^{40}$Ca${}+^{40}$Ca reaction
extracted at different center-of-mass energies. The FD potential is displayed 
with the filled circles-dotted line.
}
\label{fig:potca40ca40}
\end{center}
\end{figure}

\begin{figure}[tbhp]
\begin{center}\leavevmode
\includegraphics[width=0.85\linewidth, clip]{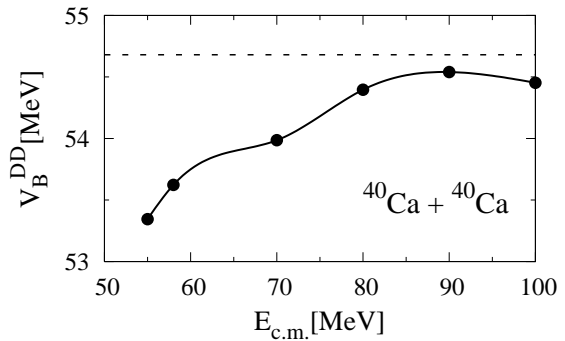}
\caption{
Barrier energy $V^{DD}_B$ for the $^{40}$Ca${}+^{40}$Ca reaction
extracted at different center-of-mass energies.
The horizontal dashed line indicates the FD reference.
}
\label{fig:ca40ca40systematics}
\end{center}
\end{figure}

\begin{figure}[bthp]
\begin{center}\leavevmode
\includegraphics[width=\linewidth, clip]{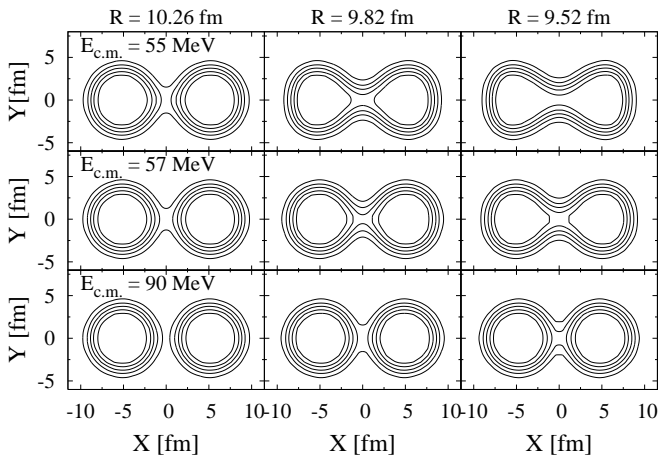}
\caption{Density profiles obtained in TDHF for different relative distances $R=10.26$ (left), $9.82$ (middle), and $9.52$~fm (right) 
for the $^{40}$Ca${}+^{40}$Ca reaction at three different center-of-mass energies: $E_{\rm c.m.}=55$, $57$, and $90$~MeV from top to bottom.
These energies corresponds to those used in Fig.~\ref{fig:ca40ca40systematics} to obtain $V^{DD}(R)$.}
\label{fig:densca40ca40}
\end{center}
\end{figure}

To illustrate the center-of-mass energy dependence of the potential, 
Figure~\ref{fig:potca40ca40} presents potentials obtained with the DD-TDHF method 
using several center-of-mass energies ranging from $E_{\rm c.m.} = 55$~MeV to $100$~MeV 
for the $^{40}$Ca${}+^{40}$Ca reaction.
Again, in the high energy limit, potentials identify with the FD case.
In addition, an increase of center-of-mass energy from $E_{\rm c.m.}=90$ to $100$~MeV
leads to identical results indicating the stability of the method as the energy increases.
In opposite, as $E_{\rm c.m.}$ decreases, potentials deduced from the DD-TDHF deviates from 
the FD case. As $E_{\rm c.m.}$ approaches the Coulomb barrier energy, a small 
change in $E_{\rm c.m.}$ significantly affects  $V^{DD}$ as illustrated by the two energies 
$E_{\rm c.m.}=55$~MeV and $57$~MeV displayed in Fig.~\ref{fig:potca40ca40}. 
In order to quantify this dependence, we have reported 
in Fig.~\ref{fig:ca40ca40systematics} values of the Coulomb barrier, denoted 
by $V^{DD}_B$, deduced from the DD-TDHF method as a function of  
center-of-mass energy.  
Again, if $E_{\rm c.m.}$ is high, $V^{DD}_B$ becomes very close to 
the FD case. As $E_{\rm c.m.}$ decreases,
$V^{DD}_B$ is more and more reduced compared to $V^{FD}_B$.
This effect, observed in all the cases considered here and called hereafter 
"dynamical barrier reduction", is a direct consequence of 
reorganization of densities in the approaching phase. 
This is clearly illustrated in Fig.~\ref{fig:densca40ca40} 
where density profiles obtained for the $^{40}$Ca${}+^{40}$Ca reaction 
at three center-of-mass energies ($E_{\rm c.m.}=55$, $57$, and $90$~MeV 
are shown from top to bottom respectively) and for specific $R$ values.  
In Fig.~\ref{fig:densca40ca40}, only the case of $E_{\rm c.m.}=90$~MeV
resembles the FD case. At lower energies, a clear deviation from the FD density profile
is observed. Considering $E_{\rm c.m.}=55$~MeV, 
as the two partners approach deformation of the two nuclei takes place. This 
deformation initiates the formation of a neck at larger relative distances compared to $E_{\rm c.m.} = 90$~MeV. 
The center-of-mass energy dependence of potential extracted with the DD-TDHF technique 
reflects the difference in the density profiles accessed 
dynamically during the mean-field evolution. Note that similar dependence is  {\it a priori} 
also expected in the DC-TDHF method~\cite{umar06,umar06c} which accounts for the dynamical deformation of the densities. 

\begin{figure}[tbhp]
\begin{center}\leavevmode
\includegraphics[width=0.85\linewidth, clip]{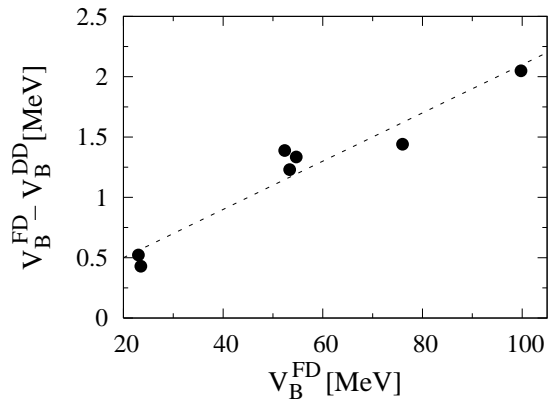}
\caption{Difference between the barrier obtained with
the FD approximation and the lowest barrier with DD-TDHF 
as a function of $V_B^{FD}$. 
In practice, the lowest barrier is obtained by using 
center-of-mass energy at or close to $V_B^{FD}$.}
\label{fig:lowvb}
\end{center}
\end{figure}

In all cases considered in this work, a reduction of the "apparent" Coulomb barrier seen by the two nuclei before fusion 
is observed compared to the FD case. This reduction could always be assigned to large density deformation close to the 
barrier. In order to quantify the magnitude of the dynamical reduction effect, we have systematically extracted the lowest 
barrier energy. This quantity (denoted by $V^{DD}_B$ (low $E_{\rm c.m.}$)), 
reported in the third column of Table~\ref{tab:vbrb}, is obtained when the center-of-mass 
energy used in DD-TDHF equals the corresponding $V^{FD}_D$. 
In Fig.~\ref{fig:lowvb}, the difference between the lowest barrier and the FD barrier is displayed as a function of 
$V^{FD}_D$. We see that the difference increases almost linearly with $V^{FD}_D$, i.e., with the initial $Z_PZ_T$.
As mentioned previously, the Coulomb barrier energy obtained within the FD approximation generally 
overestimates the Coulomb barrier energy deduced from experiments (see Table~\ref{tab:vbrb}). The discrepancy increases 
as $Z_PZ_T$ increases. Interestingly enough, the lowest energy $V^{DD}_B$ is much closer to the experimental 
observation in particular for large $Z_PZ_T$.     

\subsubsection{Critical discussion on the one-dimensional reduction: The $^{16}$O$+^{208}$Pb case}  

Previous discussions point out that various density profiles might be accessed depending on the 
center-of-mass energy used in TDHF. In macroscopic models, such a diversity in densities 
is usually accounted for by considering multidimensional collective space where deformation and/or neck 
are explicitly treated as relevant 
variables~\cite{norenberg76,hasse88,broglia91,reisdorf94,froebrich96}. 
Therefore, the energy dependence 
of the potential deduced with the DD-TDHF method should {\it a priori} be understood as different paths
in a more complex multidimensional potential energy landscape. As a consequence, one should also {\it a priori}
consider macroscopic reduction of TDHF with additional collective degrees of freedom
which might become extremely complicated.   

Here, we show that the simple one-dimensional macroscopic reduction still
contains meaningful information on the fusion process.  We consider the $^{16}$O${}+^{208}$Pb reaction 
for which extensive TDHF calculations have been performed~\cite{simenel08,umar08}. 

\begin{figure}[tbhp]
\begin{center}\leavevmode
\includegraphics[width=0.9\linewidth, clip]{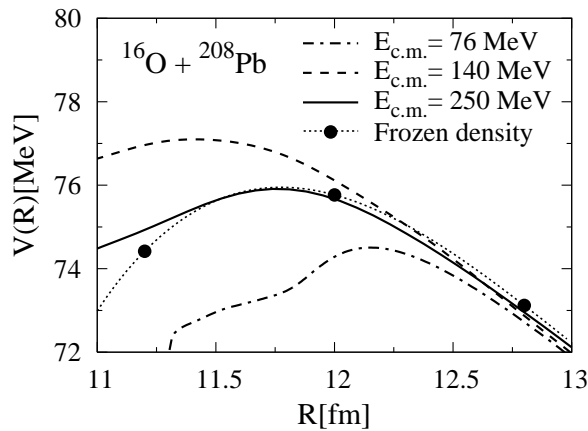}
\caption{
Potential energy obtained with the DD-TDHF method for the $^{16}$O${}+^{208}$Pb reaction
using different center-of-mass energies. The filled circles correspond to 
the FD approximation. 
}
\label{fig:poto16pb208}
\end{center}
\end{figure}

\begin{figure}[tbhp]
\begin{center}\leavevmode
\includegraphics[width=0.85\linewidth, clip]{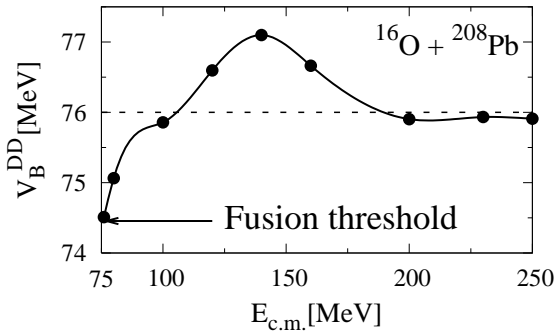} 
\caption{Coulomb barrier energy $V^{DD}_B$ obtained for the $^{16}$O${}+^{208}$Pb reaction
as a function of the center-of-mass energy. The horizontal dashed line indicates 
the FD reference while the arrow indicates the low energy fusion TDHF threshold obtained 
with TDHF calculations in Ref.~\cite{simenel08}.
}
\label{fig:poto16pb208systematics}
\end{center}
\end{figure}

Different potentials deduced for this reaction using the DD-TDHF method with different center-of-mass energies 
are displayed in Fig.~\ref{fig:poto16pb208}. A more complex energy dependence is observed in this case compared 
with the $^{40}$Ca${}+^{40}$Ca reaction displayed in Fig.~\ref{fig:potca40ca40}. In particular at intermediate
center-of-mass energies ($100$~MeV $< E_{\rm c.m.} < 200$~MeV), 
the nucleus-nucleus potential is above the FD case. This is clearly illustrated 
in Fig.~\ref{fig:poto16pb208systematics} 
where the Coulomb barrier energy $V^{DD}_B$ is shown as a function 
of $E_{\rm c.m.}$. A bump for intermediate center-of-mass energies is clearly seen. 
Note that a similar behavior is also observed in 
the $^{40}$Ca$+^{90}$Zr case indicating that the energy dependence 
might be more complex as $Z_PZ_T$ increases.
However, in view of the potential change compared to the center-of-mass energy involved (almost 
two times $V^{FD}_D$) this bump is not expected to change drastically the fusion probability obtained 
with TDHF. On opposite, when the center-of-mass energy is close to the Coulomb barrier, a small change 
in the potential will modify significantly the fusion probability. 
In the following, we will concentrate on this region.

The low energy fusion TDHF threshold which is defined as the minimal center-of-mass energy required 
to fusion in TDHF is also presented as an arrow in Fig.~\ref{fig:poto16pb208systematics}. 
A very precise value of this threshold has been obtained in Ref.~\cite{simenel08} using the same 
TDHF code with the SLy4d Skyrme effective interaction by performing a large number of TDHF
calculations and was found to be $74.45$~MeV. The lowest barrier energy obtained 
with the DD-TDHF method perfectly matches this threshold. Again, this gives additional 
confidence in this method to provide precise information on nucleus-nucleus potential extracted.

\subsection{Effect of coordinate-dependent mass}
\label{section:mu}

\begin{figure}[tbhp]
\begin{center}\leavevmode
\includegraphics[width=0.85\linewidth, clip]{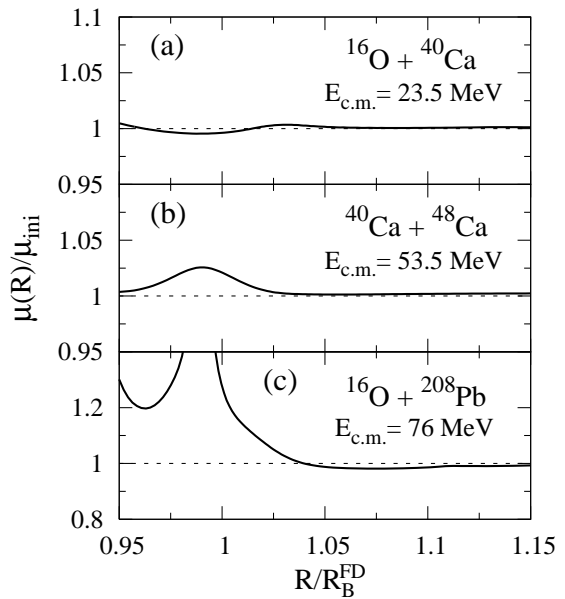} 
\caption{
Reduced mass deduced from the procedure described in Sec.~\ref{section:procedure} divided by its initial value 
$\mu_{\rm ini}$ as a function of relative distance divided by 
the Coulomb barrier radius for the FD case ($R_B^{FD}$) 
for the three mass-asymmetric reactions $^{16}$O$+^{40}$Ca (a), 
$^{40}$Ca$+^{48}$Ca (b) and $^{16}$O$+^{208}$Pb (c). 
The center-of-mass energies used are reported in each panel.
}
\label{fig:mass}
\end{center}
\end{figure}

In previous sections, the macroscopic equation~(\ref{newtonequation}) 
with neglecting the term $\frac{1}{2}\frac{d\mu(R)}{dR} \dot{R}^2$ is used as a starting point
to extract potentials from the DD-TDHF method.
This equation is {\it a priori} only valid for systems with constant reduced mass.
This condition is exactly fulfilled by the mean-field evolution for symmetric collisions.
For asymmetric reactions, dependence of the reduced mass with relative distance is possible.
This situation is illustrated in Fig.~\ref{fig:mass}, where the reduced mass estimated through 
Eq.~(\ref{eq:mu}) divided by its initial value $\mu_{\rm ini}$ is shown as a function of $R/R_B^{FD}$. 
In all cases, a deviation from the initial 
value is observed. 
In particular, this deviation might be significant for the most asymmetric case $^{16}$O$+^{208}$Pb
at small relative distances. Similar behavior has been discussed in Ref.~\cite{umar06c}.
In all cases, the center-of-mass energy used in the calculation corresponds to $V^{FD}_B$ (see table~\ref{tab:vbrb}).

\begin{figure}[tbhp]
\begin{center}\leavevmode
\includegraphics[width=0.85\linewidth, clip]{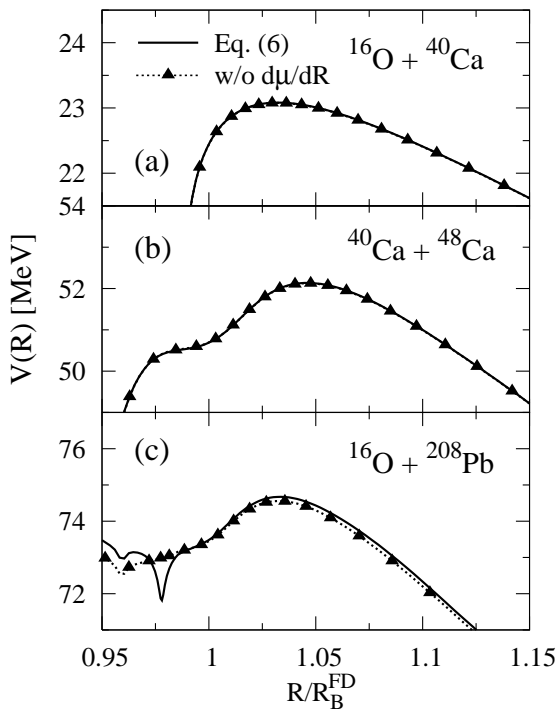} 
\caption{
Comparison of potentials extracted from Eq.~(\ref{newtonequation}) (solid line) and from 
Eq.~(\ref{newtonequation}) without the term $\frac{1}{2}\frac{d\mu(R)}{dR} \dot{R}^2$
(filled triangles-doted line)
as a function of $R/R_B^{FD}$. The center-of-mass energies used are the same as in Fig.~\ref{fig:mass}.
}
\label{fig:potcompare}
\end{center}
\end{figure}
To estimate the possible effect of the $R$-dependent reduced mass 
on the extracted potential and friction coefficient, we also 
extracted potentials and friction coefficients from Eq.~(\ref{newtonequation}) 
including the term $\frac{1}{2}\frac{d\mu(R)}{dR} \dot{R}^2$.
Using the same procedure as described in Sec.~\ref{section:procedure}, 
a new potential $V^{DD}$ can be extracted.
In Fig.~\ref{fig:potcompare}, potentials with (solid lines) and 
without (filled triangles) this term are compared for the mass-asymmetric 
$^{16}$O$+^{40}$Ca, $^{40}$Ca$+^{48}$Ca, and $^{16}$O$+^{208}$Pb reactions. 
In the first two cases, the two potentials are indistinguishable 
while in the last case a small difference between the two 
potentials is observed.   
Though we see from Fig.~\ref{fig:mass} that the reduced mass depends 
on the relative distance, Fig.~\ref{fig:potcompare} shows that 
this dependence has almost no effect on the nucleus-nucleus potential 
extracted with the DD-TDHF method.

\section{summary}
The first goal of the present paper was to  benchmark a technique, called DD-TDHF, to obtain 
nucleus-nucleus potentials and dissipation using a macroscopic reduction of mean-field 
theory based on Eq.~(\ref{newtonequation}). Several results have been obtained 
that validate the DD-TDHF technique: {\bf (i)} Used with the same conditions, the DD-TDHF method leads to potential very close to the 
DC-TDHF result~\cite{umar06} (Fig.~\ref{fig:poto16o16}). {\bf (ii)}
As expected, using high center-of-mass energies, the DD-TDHF method converges toward 
the FD approximation (Fig.~\ref{fig:potall}).
{\bf (iii)} At center-of-mass energy close to the Coulomb barrier energy, the dynamical reduction of the barrier found with the 
DD-TDHF is able to reproduce the low energy TDHF fusion threshold obtained in Ref.~\cite{simenel08} for $^{16}$O$+^{208}$Pb 
(Fig.~\ref{fig:poto16pb208systematics}).       
Nucleus-nucleus potentials obtained with the DD-TDHF method automatically incorporate dynamical effects during 
the approaching phase which could be traced back in the energy dependence of the nucleus-nucleus potential. 
This energy dependence has been systematically investigated 
for the mass symmetric reactions $^{16}$O${}+^{16}$O, 
$^{40}$Ca${}+^{40}$Ca, $^{48}$Ca${}+^{48}$Ca and
mass asymmetric systems $^{16}$O$ +^{40,48}$Ca, $^{16}$O$+^{208}$Pb, 
$^{40}$Ca${}+^{48}$Ca, $^{40}$Ca$+^{90}$Zr. For this systematic, the following aspects have been 
discussed:
\begin{itemize}
\item We show that in all reactions the minimal energy~($E^{FD}_{\rm c.m.}$)  
for which the FD potential is recovered with the DD-TDHF can always be identified. 
This energy has been systematically investigated. We have shown, that $E^{FD}_{\rm c.m.}/V^{FD}_{\rm B}$ increases as the $Z_PZ_T$ increases (Fig.~\ref{fig:ecmall}).
\item A clear energy dependence of extracted potential, due to dynamical effects which modify density profile, has been observed 
in all cases.  For systems with $Z_PZ_T \leq 400$ a continuous decrease of the apparent Coulomb barrier 
is seen as the center-of-mass energy decreases (Fig.~\ref{fig:ca40ca40systematics}) 
while in other systems ($^{16}$O$+^{208}$Pb and $^{40}$Ca$+^{90}$Zr) 
more complex energy dependence of the Coulomb barrier energy is obtained (Fig.~\ref{fig:poto16pb208systematics}).  
\item In all cases, nucleus-nucleus potential deduced from the DD-TDHF 
method varies rapidly as the center-of-mass energy
approaches the Coulomb barrier energy. Such a rapid change could be assigned to the difference 
in density profiles dynamically obtained in various TDHF calculations performed with slightly different $E_{\rm c. m.}$.
\item Dynamical effects induce a reduction of the apparent barrier compared to the FD case 
of the order $2-3 \%$ of $V^{FD}_B$ (Fig.~\ref{fig:lowvb}). While the FD Coulomb barrier generally 
overestimates the Coulomb barrier estimated experimentally, barriers including the dynamical reduction effect become 
very close to the experimental case (Table~\ref{tab:vbrb}).
\end{itemize}   
In summary, the DD-TDHF method has been successfully tested in the present work. It gives interesting 
insight in nucleus-nucleus potentials which account for dynamical effects. Mean-field calculations 
show energy dependence close to the Coulomb barrier energy.
Such energy dependence is also expected to affect the sub-barrier fusion process. 
Unfortunately, TDHF does not provide a correct description of tunneling in collective space. 
The study of sub-barrier fusion with mean-field theory would be very interesting but clearly requires 
to go beyond TDHF.
Finally, we would like to mention that the DD-TDHF technique also gives 
dissipative kernels from a fully 
microscopic theory. This aspect, which is crucial in macroscopic theory, 
will be discussed in a forthcoming article~\cite{Kou08}.   

\begin{acknowledgments}
We thank P.~Bonche for providing the 3D TDHF code and 
S.~Ayik, B.~Avez, D.~Boilley, C.~Simenel, and B.~Yilmaz  
for fruitful discussions.
One of us (K.W.) was supported by Research Fellowships of the 
Japan Society for the Promotion of Science for Young Scientists and
acknowledges GANIL for warm hospitality.
\end{acknowledgments}

\end{document}